\documentclass[aps,prl,longbibliography,twocolumn,superscriptaddress]{revtex4-2}
\usepackage{amsmath,amssymb,bm,graphicx,color,gensymb,bbold,hyperref,keyval,url,latexsym}
\usepackage[dvipsnames,svgnames,table]{xcolor}
\usepackage{enumitem}
\usepackage{comment}

\hypersetup{pdfstartview={FitH}, colorlinks=true, linkcolor=NavyBlue, citecolor=blue, filecolor=NavyBlue, urlcolor=NavyBlue}

\newcommand{\bk}{\bm{k}}

\begin{document}

\title{Repulsion-Driven $p - i p$ Superconductivity in a Single Valley Revealed by DMRG}
\author{Cesar A. Gallegos}
\author{Omid Tavakol}
\author{Christopher Yang}
\author{Steven R. White}
\author{Thomas Scaffidi}
\affiliation{Department of Physics and Astronomy, University of California, Irvine, California 92697, USA}
\date{\today}
\begin{abstract}
We demonstrate repulsion-driven topological superconductivity of a single species of Dirac fermions, motivated by valley-polarized phases observed in two-dimensional materials such as rhombohedral graphene.
Using density-matrix renormalization group calculations on the Qi--Wu--Zhang lattice model on cylinders of width up to eight, we find a robust phase of spinless chiral $p$-wave superconductivity.
Pairing already starts at low doping and thus occurs on a small, nearly isotropic Fermi pocket, and is therefore not tied to the particular details or anisotropy of the dispersion but instead seems tied to the non-trivial quantum geometry of the bands. 
In fact, the winding of the superconductor order parameter is opposite (``$p-ip$'') to that of the anomalous Hall metal which forms the parent state, in agreement with expectations recently derived from weak coupling calculations.
At larger doping, we find that a pair-density-wave component develops alongside zero-momentum pairing.
Our work shows that the combination of strong repulsion, absence of time-reversal in the parent state, and non-trivial quantum geometry form a promising platform to realize topological superconductivity.
\end{abstract}
\maketitle

{\it Introduction.}---%
The momentum-space structure of Bloch wave functions can profoundly influence superconductivity. 
The canonical example is the surface of a three-dimensional topological insulator, which realizes a single Dirac cone. 
In the Fu--Kane construction, conventional $s$-wave pairing projected onto this surface state inherits the winding of the Dirac spinor and becomes topologically non-trivial~\cite{FuKane2008, Santos2010, QiHughesZhang2010, FuBerg2010, PotterLee2011, LeeHazraRanderiaTrivedi2019, QiZhang2011, BlackSchafferBalatsky2013, DasSarmaLi2013, BrydonDasSarmaHuiSau2014, XuLianTangQiZhang2016, ZhangEtAl2018, WangEtAl2018}. 
Beyond surface states, a single Dirac fermion can also arise in strictly two-dimensional materials when time-reversal symmetry is broken. 
An experimentally relevant setting is the quarter-metal phase of rhombohedral $N$-layer graphene, where superconductivity has been observed near spin- and valley-polarized metallic states~\cite{ZhouEtAl2021, ChoiEtAl2025, HanEtAl2025, CeaPantaleonPhongGuinea2022, YouVishwanath2022, ParraMartinezEtAl2025, FermiologyRhombohedral2026, GhazaryanHolderSerbynBerg2021, GhazaryanHolderBergSerbyn2023, ChouZhuDasSarma2025, LongJimenoPozoSainzCruzPantaleonGuinea2024}.
In an idealized continuum description, these systems correspond to higher-winding $N$-Dirac models, whose Bloch spinors wind $N$ times around the Fermi surface~\cite{KoshinoMcCann2009}.
Superconductivity emerging from an anomalous Hall metal has also been reported in twisted bilayer MoTe$_2$~\cite{XuEtAl2025}.
Together, these settings motivate the study of pairing on a single nondegenerate Fermi surface with nontrivial Bloch-spinor winding. 

In this work, we focus on intrinsic superconductivity that emerges spontaneously from repulsive interactions. 
The Kohn--Luttinger mechanism allows repulsion to generate attractive pairing channels, but the resulting transition temperatures are typically very low for simple, nearly parabolic bands with trivial orbital content, particularly in two dimensions~\cite{ChubukovLu1992, Chubukov1993}. 
Achieving an appreciable critical temperature therefore relies on favorable band-structure features, such as specific Fermi-surface shapes or proximity to a van Hove singularity~\cite{KohnLuttinger1965, BaranovKagan1992, ChubukovLu1992, Chubukov1993, Hlubina1999, RaghuKivelsonScalapino2010, RaghuKivelson2011, Gonzalez2008, NandkishoreLevitovChubukov2012, NandkishoreThomaleChubukov2014, KaganValkovMitskanKorovushkin2014, MaitiChubukov2013, ChubukovKivelson2017, Scaffidi2017, PhysRevB.94.085106, PhysRevB.98.224515, GonzalezStauber2019, ChichinadzeClassenChubukov2020, PhysRevB.107.014505}. 
Recent work has identified nontrivial quantum geometry as a distinct route to enabling repulsion-driven superconductivity: the momentum-dependent texture of Bloch spinors enters the projected interaction through band form factors and can generate pairing even on a small, nearly isotropic Fermi surface~\cite{KitamuraDaidoYanase2024, ShavitAlicea2025, JahinLin2025, GeierDavydovaFu2026, MayMannHelbigDevakul2025, PatriFranz2025, MurshedDasRoy2025, DongLee2025, TavakolScaffidi2026}.

In this work, we address the question of whether this quantum-geometry-enabled electronic mechanism for superconductivity survives to intermediate and strong coupling, which is often the experimentally relevant regime and has the potential of realizing higher critical temperatures.
We consider the minimal case of a single Dirac fermion, realized in a two-dimensional lattice model with broken time-reversal symmetry (also known as ``Wilson fermions''). 
Weak-coupling calculations predict a $p-ip$ superconducting instability for small $U$, with winding opposite to that of the anomalous Hall metal above $T_c$~\cite{TavakolScaffidi2026} (see also Ref.~\cite{ShavitAlicea2025}). 
To study this system in the strong repulsion regime, we use density-matrix renormalization group (DMRG) calculations on the Qi--Wu--Zhang (QWZ) model, which can be tuned to a lattice realization of a single Dirac fermion~\cite{QiWuZhang2006}, augmented with on-site repulsion. 
On cylinders of width up to eight and at interaction strength $U=6$, well beyond the weak-coupling regime, we find strong evidence for  $p-ip$ superconductivity upon doping. 
Our results thus provide evidence that a valley- and spin-polarized metal provides a particularly promising platform for realizing a spinless chiral $p$-wave superconductor, the ``holy grail'' of topological superconductivity.

We note that other non-perturbative studies have investigated repulsion-driven pairing and superconductivity in complementary settings, including multivalley systems with sublattice polarization~\cite{HeEtAl2023}, narrow topological bands~\cite{PhysRevB.106.035421, SahayEtAl2024, GuerciAbouelkomsanFu2025}, doped Chern ferromagnets~\cite{gonzalves2026}, and models hosting fractional Chern insulators or chiral spin liquids~\cite{WangZaletel2025, Divic2025, Chen2026, Soejima2026}.

\begin{figure}[t!]
\includegraphics[width=\linewidth]{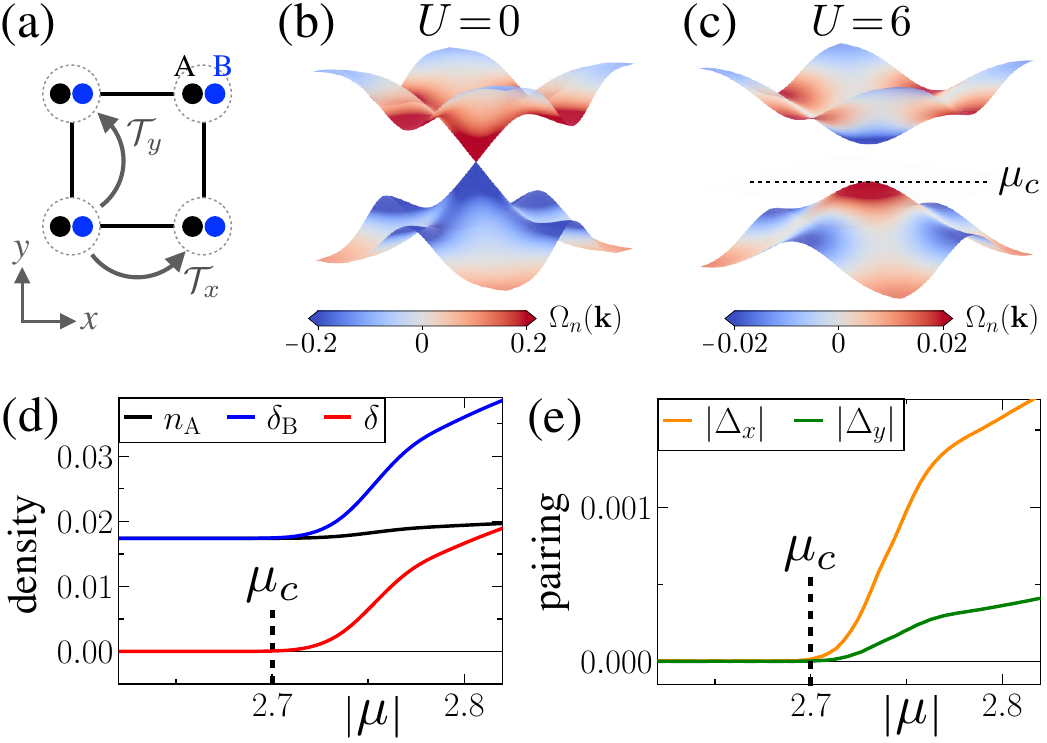}
\vskip -0.3cm
\caption{(a) Illustration of the QWZ model with two orbitals on a square lattice and (b) its non-interacting band structure with a Dirac point at the center of the Brillouin zone. 
Color indicates Berry curvature.
(c) In the strong $U$ limit, the Hartree bands open an effective mass gap $\propto U$ at the Dirac point (the band gap has been artificially reduced for ease of plotting). 
The superconducting phase emerges upon doping (we dope with holes without loss of generality, for $\mu < \mu_c$).
(d), (e) DMRG scan on an $84\!\times\!8$ cylinder for $U\!=\!6$, $\mathcal{B}\!=\!0.5$, with a linearly decreasing $\mu$ from $\mu\!=\!-2.6$ to $\mu\!=\!-2.9$, and $\mu_c \!\sim\! -2.7$ indicating the lower edge of the charge gap.
(d) Average occupation $n_{\rm A}$ and hole-doping densities $\delta_B \!=\!1\!-\!n_{\rm B}$ and $\delta \!=\! 1\!-\!n_{\rm A}\!-\!n_{\rm B}$.
(e) Column-averaged nearest-neighbor pairing strength in $x$ and $y$ directions, $|\Delta_{x(y)}|$.}
\label{fig:Scan_INS-SC}
\vskip -0.5cm
\end{figure}

{\it Model and method.}---%
We consider an interacting version of the massless ($m\!=\!0$) QWZ model~\cite{QiWuZhang2006} on a square lattice with orbitals A and B, described by the Hamiltonian $H\!=\!H_0 + H_U$, where
\begin{equation}
H_0 = \sum_{i} (m+2\mathcal{B}) \mathbf{c}_{i}^{\dagger}  \sigma_z \mathbf{c}_{i}  + \big(\mathbf{c}_{i+\hat{x}}^{\dagger} \mathcal T_x \mathbf{c}_{i} + \mathbf{c}_{i+\hat{y}}^{\dagger} \mathcal T_y \mathbf{c}_{i}  + {\rm H.c.}\big),
\label{eq:QWZ}
\end{equation}
with $\mathcal T_{x(y)} \!=\! -\frac{\mathcal B}{2} \sigma_z \mp \frac{iv}{2} \sigma_{y(x)}$, and
\begin{align}
H_U&=U\sum_{i}n_{i, \rm A}n_{i, \rm B}, \quad U > 0,
\end{align}
is an on-site repulsive interaction between the orbitals. 
Here, $n_{i,\alpha}\!=\!c_{i,\alpha}^{\dagger}c_{i,\alpha}$ is the density on orbital $\alpha\!=\!\rm A,\rm B$, $\mathbf{c}_i\! =\! (c_{i,\rm A}, c_{i,\rm B})$, $\sigma$ are the Pauli matrices, and $\hat{x}$ and $\hat{y}$ are unit-vectors in the $x$- and $y$-directions; see Fig.~\ref{fig:Scan_INS-SC}(a).

The QWZ model is the spin-polarized version of the Bernevig-Hughes-Zhang (BHZ) model describing the spin Hall effect, in which case the two orbitals correspond to the $s$ and $p$ orbitals of a semiconductor~\cite{QiWuZhang2006, BernevigHughesZhang2006} (see Refs.~\cite{WangDaiXie2012, BudichTrauzettelSangiovanni2013, AmaricciEtAl2015, GilardoniEtAl2022, MiyakoshiOhta2013, SoniEtAl2024, FavataEtAl2025} for studies of interacting versions of BHZ). 
It can also be realized with cold atoms~\cite{PhysRevResearch.5.L012006}.

For $U\!=\!0$, $m\!=\!0$ and $\mathcal{B}>0$, this model provides a minimal lattice realization of a single massless Dirac fermion at the transition between a trivial ($m>0$) insulator and a Chern ($m<0$) insulator. 
The time-reversal-breaking $\mathcal{B}$ term acts as a momentum-dependent Wilson mass, gapping the additional lattice Dirac points while leaving a single gapless cone at $\Gamma\!=\!(0,0)$. 
This construction implements the Wilson-fermion regularization widely used in lattice QCD~\cite{PhysRevD.10.2445}. 
The Bloch Hamiltonian reads
\begin{equation}
\mathcal{H}(\bk)
= v\sin k_y\sigma_x-v\sin k_x\sigma_y
+\mathcal{B}(2-\cos k_x-\cos k_y)\sigma_z,
\label{eq}
\end{equation}
which near $\Gamma$ reduces to $\mathcal{H}(\bk)\simeq vk_y\sigma_x-vk_x\sigma_y+\frac{1}{2}\mathcal{B}k^2\sigma_z$; see Fig.~\ref{fig:Scan_INS-SC}(b). 
The Wilson term also produces a nonzero orbital polarization, $n_{\rm B}-n_{\rm A}\neq0$, whose sign follows that of $\mathcal{B}$. 
Here $n_\alpha\equiv\langle n_{i,\alpha}\rangle$ denotes the average occupation of orbital $\alpha={\rm A},{\rm B}$, with $n_{\rm A}+n_{\rm B}=1$ at half filling.

\begin{figure*}[ht!]
\includegraphics[width=.9\linewidth]{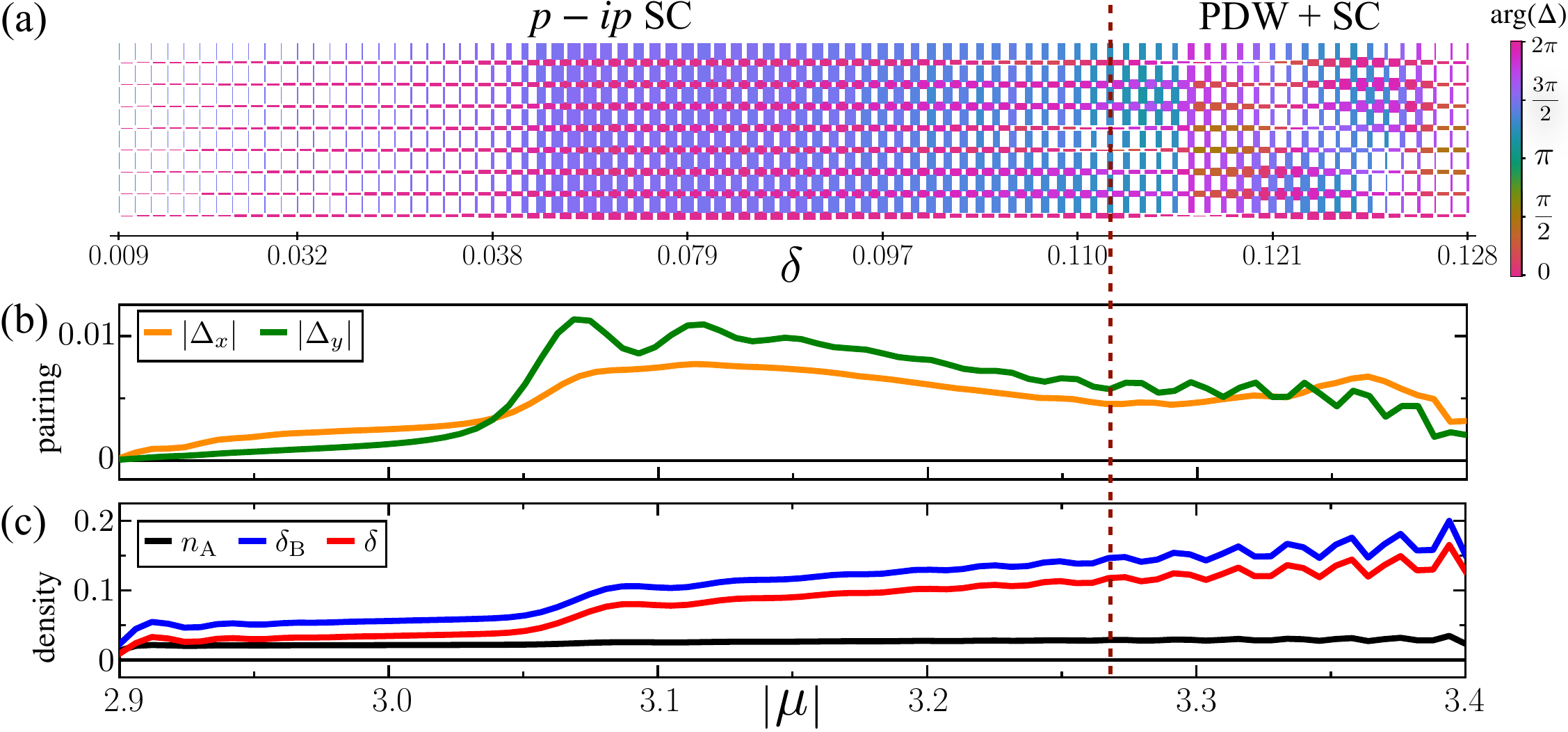}
\vskip -0.3cm
\caption{DMRG scan on an $84\!\times\!8$ cylinder for $U\!=\!6$ and $\mathcal{B}\!=\!0.5$, with the chemical potential decreasing linearly along the cylinder from $\mu\!=\!-2.9$ to $\mu\!=\!-3.4$ and a $10^{-4}$ pair field with $p-ip$ winding applied across the cylinder.
(a) DMRG cylinder showing the $p-ip$ superconducting (SC) phase at lower doping and the phase with coexisting pair-density-wave order and uniform superconductivity (PDW+SC) at higher doping.
The bond thickness is proportional to the nearest-neighbor (NN) pairing strength on the B orbitals, $|\Delta|\!=\!|\langle c_{i,{\rm B}}c_{j,{\rm B}}\rangle|$, while the color indicates the phase $\arg(\Delta)$ relative to the bottom-row horizontal bond of each column. 
Column-averaged (b) NN pairing strengths $|\Delta_x|$ and $|\Delta_y|$ in the $x$ and $y$ directions, respectively, and (c) A-orbital occupation $n_{\rm A}$, B-orbital hole density $\delta_{\rm B}$, and total hole density $\delta$.
For clarity, we use $\delta$, extracted from the density profile in (c), as the horizontal-axis coordinate in panel (a).}
\label{fig:Scan_SC-PDW}
\vskip -0.5cm
\end{figure*}

At half filling, turning on $U>0$ opens a charge gap and further enhances the orbital polarization, see Fig.~\ref{fig:Scan_INS-SC}(c). 
Within Hartree theory, this is described by an effective mass term, $\mathcal{H}(\bk)\to\mathcal{H}(\bk)+m_{\mathrm{eff}}\sigma_z$, with $m_{\mathrm{eff}}=U(n_{\rm B}-n_{\rm A})/2$. 
In the strongly polarized regime, $m_{\mathrm{eff}}\simeq(U/2)\mathrm{sign}(\mathcal{B})$; see Supplemental Material (SM)~\cite{SupplementalMaterials}. 
Since $\mathcal{B}$ and $m_{\mathrm{eff}}$ have the same sign, the resulting Hartree bands are topologically trivial, consistent with an orbital-polarized trivial insulator in the large-$U$ limit at half-filling~\cite{ZieglerEtAl2022AnnPhys,ZieglerEtAl2022PRR}. 
These bands nevertheless retain locally nonzero Berry curvature, despite their vanishing Chern number; see Figs.~\ref{fig:Scan_INS-SC} (b) and (c). 

We now use DMRG to study this model away from half-filling.
Throughout this work, we set $v\!=\!1$, $\mathcal{B}\!=\!0.5$ and $U\!=\!6$, and focus on hole-doping without loss of generality since the problem is particle-hole symmetric.
We perform DMRG calculations on cylinders of size $N_x\!\times\!N_y$ with periodicity in the transverse $y$-direction and open boundary conditions in the longitudinal $x$-direction.
Our analysis utilizes both DMRG ``scans'', in which a parameter of the Hamiltonian varies along the $x$ direction, and ``non-scan'' calculations, in which all model parameters are fixed across the cylinder. 
DMRG scans help identify phase boundaries, near which competing states are close in energy and non-scans may become trapped in metastable states. 
We used sufficiently slow parameter ramps to ensure extended regions of the cylinder lay well within each phase, helping stabilize the corresponding local order. 
As a check, we confirmed that non-scan simulations gave very similar results at matching parameter values.

We work in the grand canonical ensemble for $U(1)$ particle-number symmetry, which allows finite-bond-dimension DMRG to select a $U(1)$-symmetry-broken state, thus mimicking the thermodynamic limit and providing direct access to the local pair amplitude $\langle c_i c_j\rangle$.
Although not enforcing $U(1)$ symmetry increases the computational cost of DMRG at fixed bond dimension, superconductivity is diagnosed with local pairing observables that are converged at much smaller bond dimensions than pair-pair correlations in $U(1)$-conserving calculations~\cite{ttprimej, Chen2025}.
A bond dimension of $600$ is typically sufficient to achieve convergence, with truncation errors of $\mathcal{O}(10^{-5})$.
See End Matter for more details on the DMRG calculations.

\begin{figure}[t!]
\includegraphics[width=\linewidth]{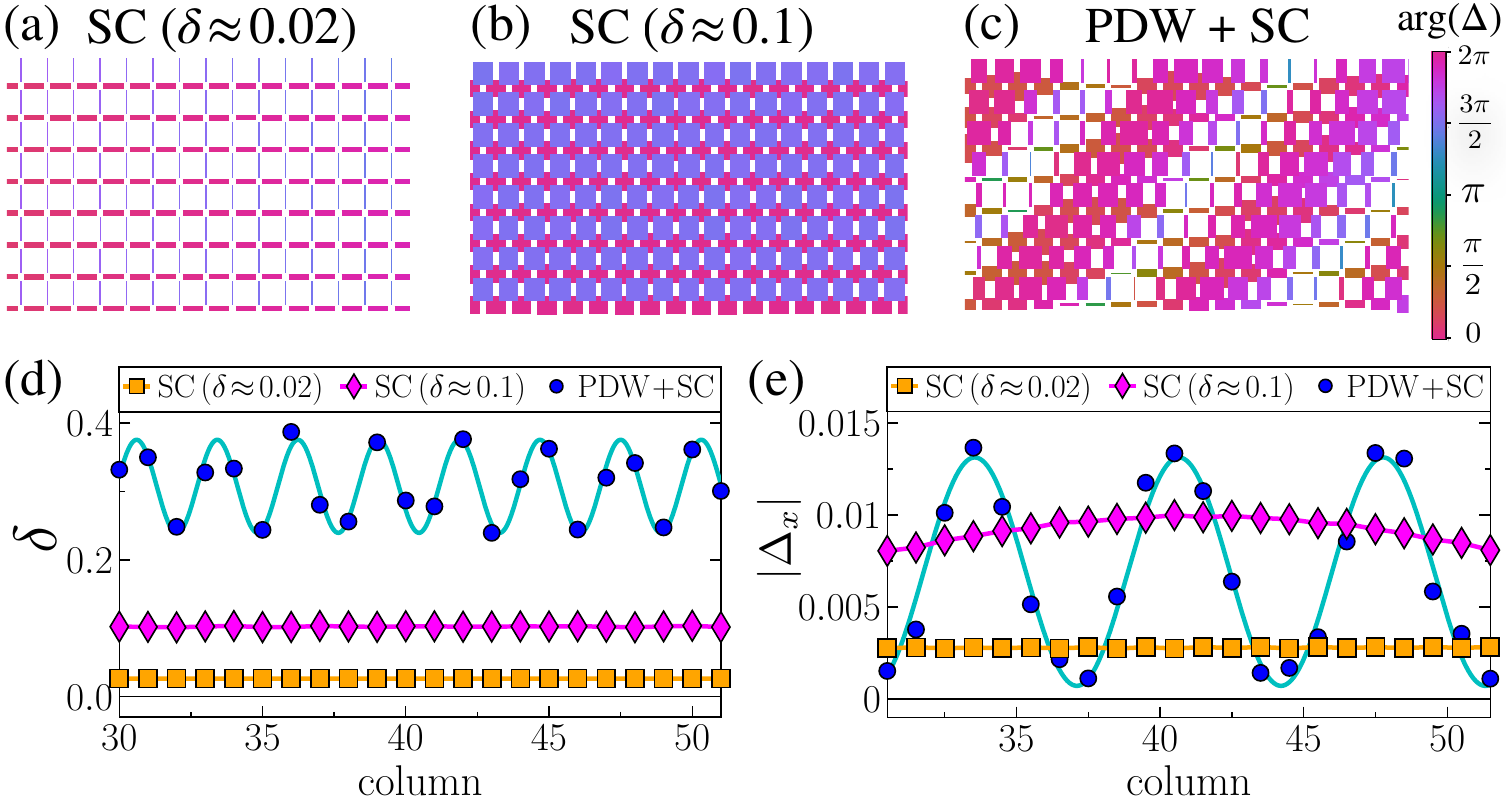}
\vskip -0.2cm
\caption{DMRG non-scan results on $84\!\times\!8$ cylinders for $U\!=\!6$, $\mathcal{B}\!=\!0.5$, and a $10^{-4}$ pair field with $p-ip$ winding applied across the cylinder. 
The DMRG cylinders on the (a), (b) SC phase at $\delta\!\approx\!0.02$ $(\mu\!=\!-2.9)$ and $\delta\!\approx\!0.1$ $(\mu\!=\!-3.2)$, respectively, and the (c) PDW+SC phase at $\delta\!\approx\!0.3$ $(\mu\!=\!-3.8)$. 
Bond thickness is proportional to the nearest-neighbor pairing strength $|\Delta|\!=\!|\langle c_{i,{\rm B}}c_{j,{\rm B}}\rangle|$, and bond color indicates the phase $\arg(\Delta)$ measured relative to a horizontal bond in the center. 
The (d) total hole density  $\delta$ and (e) pairing strength in the $x$ direction, $|\Delta_x|$, along a single row. 
The light-blue curves are independent fits of the PDW+SC data to a constant plus a cosine. 
Only the central region of each cylinder is shown.}
\label{fig:NonScanCylinder}
\vskip -0.4cm
\end{figure}

\begin{figure}[t!]
\includegraphics[width=\linewidth]{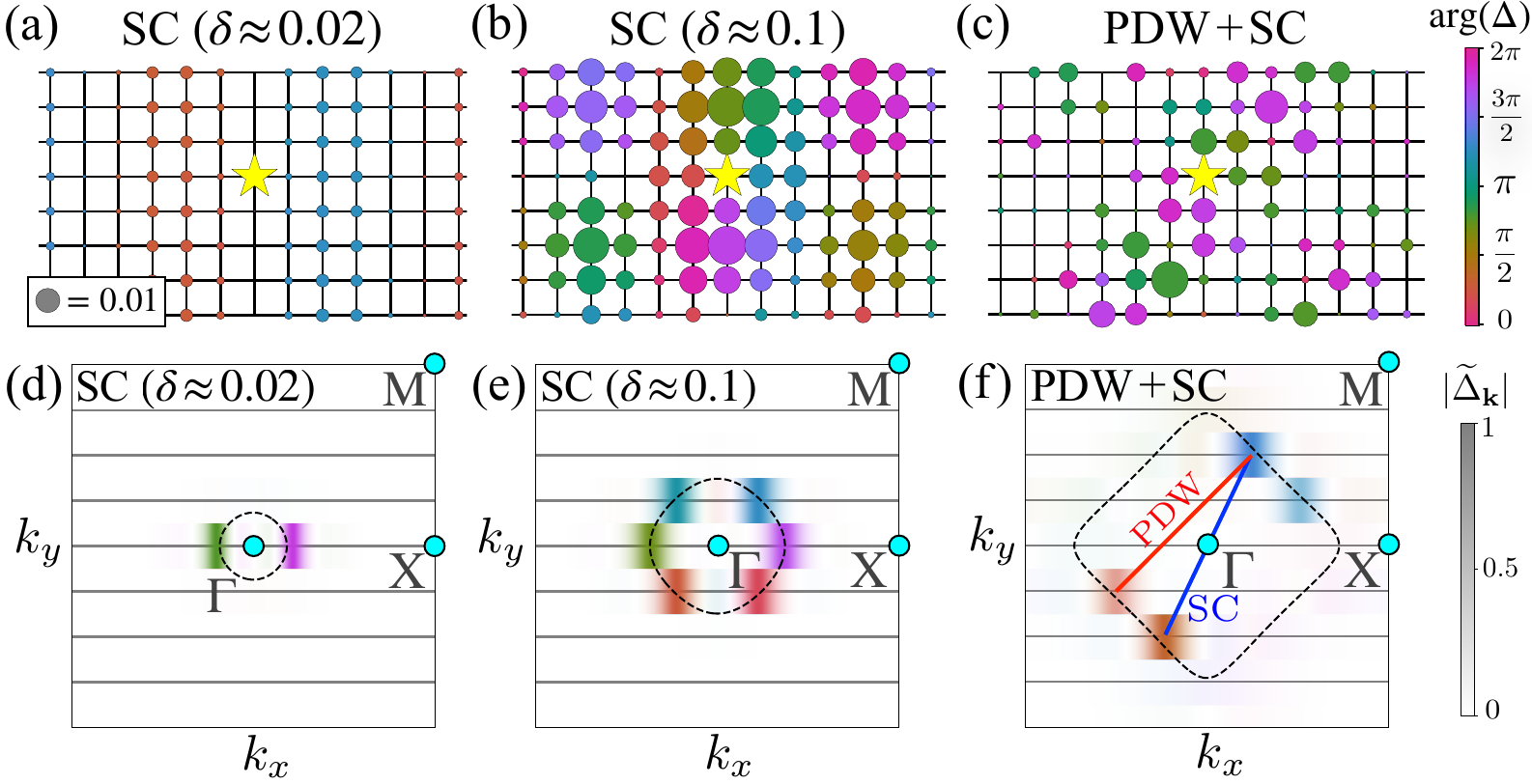}
\vskip -0.2cm
\caption{Same DMRG non-scans as in Fig.~\ref{fig:NonScanCylinder}. 
(a)--(c) The real-space pair amplitude $\langle c_{0}c_i\rangle$, measured relative to the central site marked by a star, with symbol size and color indicating its magnitude and phase, respectively. 
(d)--(f) The magnitude (gray scale) and phase (color) of the normalized zero-momentum pair amplitude $\widetilde{\Delta}_{\bf k}\!=\! \Delta_{\bf k}/\max(|\Delta_{\bf k}|)$, with $\Delta_{\bf k}\!=\!\langle c_{\bf k}c_{-\bf k}\rangle$. 
The blue and red lines in (f) connect the momenta involved in uniform SC and finite-momentum PDW pairing, respectively. 
The dashed contours indicate the Fermi surfaces obtained from the mean field band structure with an effective mass $m_{\rm eff}=U/2$ and where the doping was matched to the average one observed in DMRG. 
The horizontal gray lines indicate the discrete $k_y$ values allowed on the width-$8$ cylinder.}
\label{fig:Pairing}
\vskip -0.5cm
\end{figure}

{\it $p-ip$ SC.}---%
The central result of this work is the appearance of a $p-ip$ superconductor when doping the orbital-polarized insulator.
We analyze it based on three separate calculations: (1) a low-doping scan showing the transition from the orbital insulator to the doped regime with SC correlations (Fig.~\ref{fig:Scan_INS-SC}), (2) a wider-range chemical potential scan showing the SC phase extending from zero doping to $\delta_c \simeq 0.112$, above which a pair-density-wave also appears (Fig.~\ref{fig:Scan_SC-PDW}), and (3) fixed-$\mu$ calculations at representative doping values (Figs.~\ref{fig:NonScanCylinder} and \ref{fig:Pairing}) that show the real and momentum space structure of the gap in each phase.

Figures~\ref{fig:Scan_INS-SC}(d) and (e) show a DMRG chemical-potential scan on an $84\times8$ cylinder, with $\mu$ decreasing linearly from $-2.6$ to $-2.9$~\footnote{The chemical potential includes an offset of $-U/2$, placing the particle-hole-symmetric point at $\mu=0$.}.
The sharp increase in hole density, $\delta=1-n_{\rm A}-n_{\rm B}$, locates the upper edge of the lower band at $\mu_c\simeq-2.7$. 
For $\mu>\mu_c$, the system is insulating, with $n_{\rm A}\approx0.02$ and $n_{\rm B}\approx0.98$. Below $\mu_c$, holes enter the predominantly B-orbital valence band, mainly reducing $n_{\rm B}$.
Nonzero pair amplitudes, $\Delta_{i\alpha,j\beta}=\langle c_{i,\alpha}c_{j,\beta}\rangle$, appear at the onset of hole doping, suggesting pairing at very low doping. 
Pairing is strongest between B orbitals, which we focus on below, omitting the orbital index when convenient. 
Figure~\ref{fig:Scan_INS-SC}(e) shows the corresponding nearest-neighbor amplitudes, $\Delta_{x(y)}\equiv\langle c_{i,{\rm B}}c_{i+\hat{x}(\hat{y}),{\rm B}}\rangle$, along the $x$ and $y$ directions.

To follow the evolution of this paired state, we extend the chemical-potential scan to larger doping, with $\mu$ decreasing linearly from $-2.9$ to $-3.4$; see Fig.~\ref{fig:Scan_SC-PDW}.
This wider scan reveals the $p-ip$ SC phase extends to the entire region $0<\delta<\delta_c\simeq0.112$, with $\arg(\Delta_y/\Delta_x)=-\pi/2$ throughout that region.
For $\delta > \delta_c$, the pairing amplitude develops real space oscillations in both its magnitude and phase [best observed in Fig.~\ref{fig:Scan_SC-PDW} (a)], which can be understood as the appearance of a pair-density wave order (PDW) in addition to SC; we defer the discussion of this phase for later.

Within the SC phase, Fig.~\ref{fig:Scan_SC-PDW} shows two subregions separated by a rapid increase in both hole density and pairing strength near $\delta\simeq0.038$, accompanied by a crossing of the $|\Delta_x|$ and $|\Delta_y|$ curves.
The presence of two subregions can be understood from the discrete transverse momenta imposed by the finite cylinder width.
At low doping ($\delta<0.038$), the Fermi surface intersects only the $k_y=0$ channel, giving rise to quasi-1D behavior.
At higher doping ($\delta>0.038$), it also intersects the $k_y=\pm2\pi/N_y$ channels, which appears sufficient to recover a 2D regime representative of the thermodynamic limit [see Figs.~\ref{fig:Pairing}(d) and (e), and End Matter].
The quasi-1D region is expected to range from $\delta=0$ to $\delta=2/N_y^2$ and thus vanish as $N_y\to\infty$, consistently with the trend we observed between width-6 and width-8 cylinders (see SM~\cite{SupplementalMaterials} for the width-6 data).

We therefore focus on the range $0.038<\delta<0.112$ as representative of the $p-ip$ SC phase in the thermodynamic limit.
A typical non-scan calculation in this range, at $\delta=0.1$ ($\mu=-3.2$), shows comparable nearest-neighbor pairing strengths along $x$ and $y$, $|\Delta_x|\sim|\Delta_y|$ [Fig.~\ref{fig:NonScanCylinder}(b)].
Both the hole density and pairing amplitude are approximately uniform in the bulk, and $|\Delta| \sim 0.01$ far exceeds the pairing field strength of $10^{-4}$ [Figs.~\ref{fig:NonScanCylinder}(d) and (e), and see End Matter for more details on the DMRG].

\begin{figure}[t!]
\includegraphics[width=\linewidth]{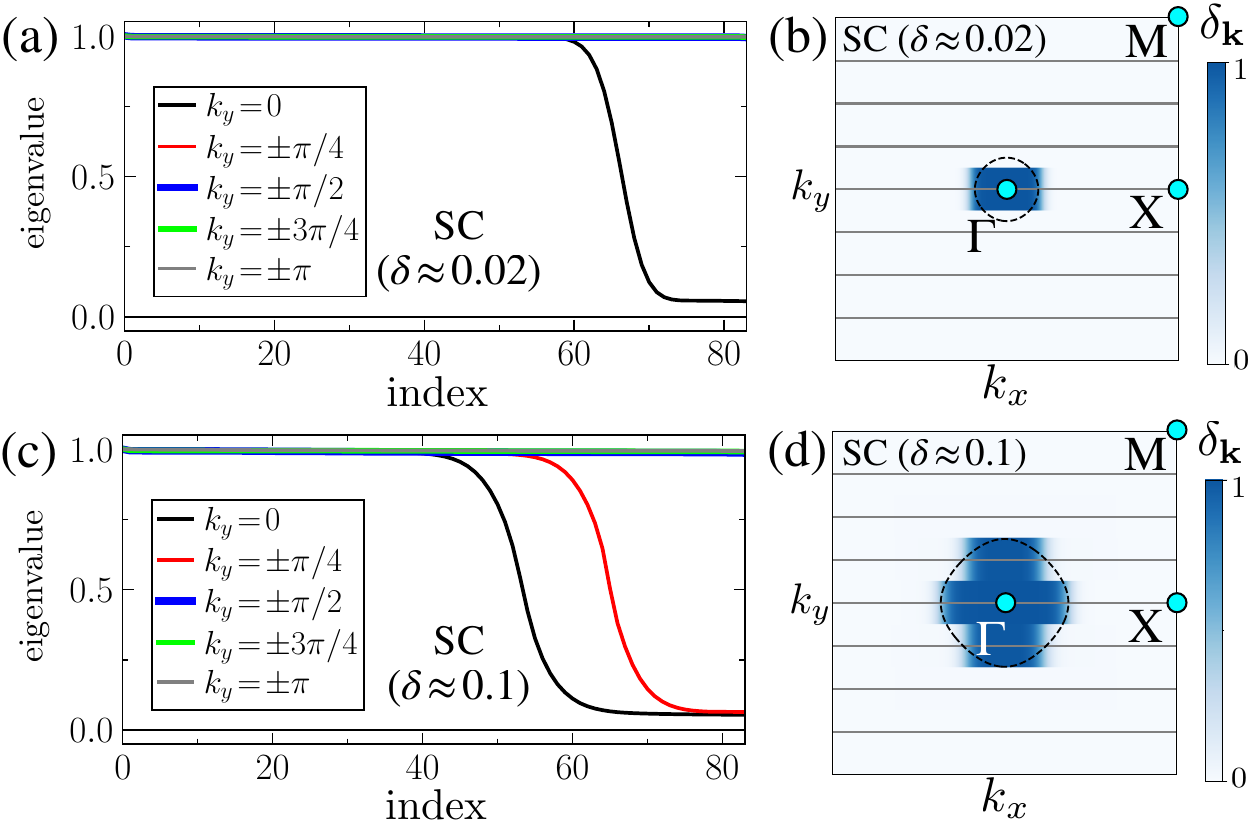}
\vskip -0.2cm
\caption{Same DMRG non-scans as in the main text, at doping $\delta\approx0.02$ and $\delta \approx 0.1$. 
Sorted eigenvalues of the one-body reduced density matrix in each $k_y$ sector are shown in (a) and (c), and the corresponding hole momentum distributions $\delta_{\bf k}$ in (b) and (d). 
The dashed contours in (b) and (d) show the Fermi surfaces obtained from the noninteracting model with the effective mass $m_{\rm eff}\!=\!U/2$ and the horizontal gray lines indicate the discrete $k_y$ values allowed on the width-$8$ cylinder.}
\label{fig:1RDM_Nk}
\end{figure}

We also extract the ``Cooper pair wavefunction'' both in real-space through superconducting correlations $\langle c_0c_i\rangle$ with $c_0$ the central site [Fig.~\ref{fig:Pairing}(b)], and in momentum space through Fourier transform [Fig.~\ref{fig:Pairing}(e)].
The $p-ip$ winding is clearly observed in both real and momentum space.
The pairing amplitude $\Delta_{\bf k}$ is sharply concentrated near the Fermi surface, with corresponding extended oscillations in real space.
Together with the sharp variation of $n({\bf k})$ across the Fermi surface (Fig.~\ref{fig:1RDM_Nk} in the End Matter), this is consistent with BCS-like weak pairing.
Combined with the $p-ip$ phase winding, these features provide strong evidence that the SC state realizes the topological weak-pairing phase of Read and Green~\cite{ReadGreen2000}.

{\it Coexisting PDW and uniform SC.}---%
Above $\delta_c\simeq0.112$, the scan in Fig.~\ref{fig:Scan_SC-PDW} shows the onset of spatial modulations in the pairing amplitude~\footnote{The hole density also oscillates, but its modulations are weaker than those of the pairing amplitude [Fig.~\ref{fig:NonScanCylinder}(d)]. Although composite order parameters of PDW+SC can generate density oscillations, these may be difficult to distinguish from Friedel oscillations originating at the edges, which may decay only as a power law along the nodal directions.}.
A non-scan calculation at $\delta\approx0.3$ ($\mu=-3.8$) confirms a stripe pattern along one of the two diagonal directions [Fig.~\ref{fig:NonScanCylinder}(c)], also visible in the real-space pair wavefunction [Fig.~\ref{fig:Pairing}(c)].
The pairing strengths $|\Delta_{x,y}|$ oscillate about a nonzero background [Fig.~\ref{fig:NonScanCylinder}(e)], consistent with coexisting PDW and uniform SC components.

\begin{figure}[t!]
\includegraphics[width=\linewidth]{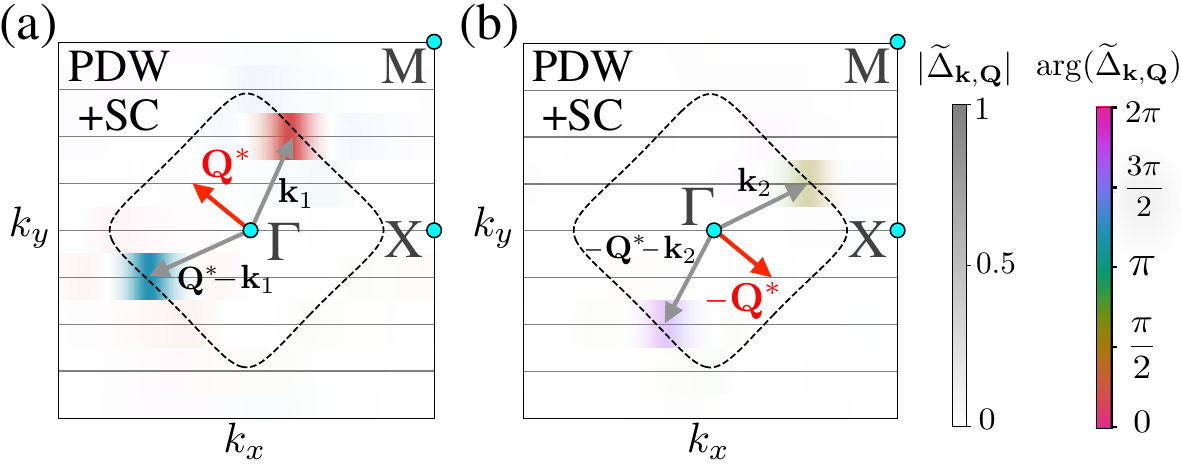}
\vskip -0.2cm
\caption{Finite-total-momentum pairing in the PDW+SC phase, for the same DMRG non-scan as in Fig.~\ref{fig:Pairing}(f).
The magnitude (gray scale) and phase (color code) of 
$\widetilde{\Delta}_{{\bf k},{\bf Q}}\!=\! \Delta_{{\bf k},{\bf Q}}/\max(|\Delta_{{\bf k},{\bf Q}}|)$, with $\Delta_{{\bf k},{\bf Q}}=\langle c_{{\bf k}}c_{{\bf Q}-{\bf k}}\rangle$ are shown as functions of ${\bf k}$ at fixed total momenta (a) ${\bf Q}^*=(-2\pi/7,\pi/4)$ and (b) $-{\bf Q}^*$.
The magnitude $\max_{{\bf k}}|\Delta_{{\bf k},{\bf Q}^*}|$ is approximately half the maximum magnitude of the zero-momentum pair amplitude in Fig.~\ref{fig:Pairing}(f).
The two dominant peaks identify the constituent momenta of a Cooper pair.
The gray arrows indicate these individual momenta, while the red arrow indicates their sum, $\pm {\bf Q}^*$.}
\label{fig:DeltaQ}
\vskip -0.5cm
\end{figure}

To resolve these components, we Fourier transform the real-space pair amplitudes to obtain $\Delta_{{\bf k},{\bf Q}}=\langle c_{{\bf k}}c_{{\bf Q}-{\bf k}}\rangle$.
We first analyze the ${\bf Q}=0$ component, plotted in Fig.~\ref{fig:Pairing}(f).
At this doping, the Fermi surface is approximately square, and $\Delta_{{\bf k},0}$ exhibits four sharp peaks at Fermi-surface momenta on one pair of opposite edges, with much weaker pairing on the other pair.
This is consistent with $p_x\pm p_y$ order.
For these four Fermi-surface momenta, it is natural to consider not only zero-momentum pairing between $(k_x,k_y)$ and $(-k_x,-k_y)$, but also finite-momentum pairing between $(k_x,k_y)$ and $(-k_y,-k_x)$.
The blue and red lines in Fig.~\ref{fig:Pairing}(f) illustrate the momenta paired in these two channels, respectively.
Indeed, certain finite-${\bf Q}$ components of $\Delta_{{\bf k},{\bf Q}}$, negligible in the uniform SC phase, become comparable in magnitude to the zero-momentum component in the PDW+SC phase.
The strongest occur at ${\bf Q}=\pm{\bf Q}^*$, with ${\bf Q}^*=(-2\pi/7,\pi/4)$, approximately along the $[1\bar{1}0]$ direction.
(The slight deviation from this direction reflects the discretization of $k_y$.)
Like the uniform SC component, the PDW component has $p_x+p_y$ symmetry in the relative coordinate [see Fig.~\ref{fig:DeltaQ} in the End Matter for plots of $\Delta_{{\bf k},\pm{\bf Q}^*}$].

We now describe the center-of-mass dependence by retaining the three dominant components, with amplitudes $\Delta(0)=|\Delta_{\mathrm{SC}}|e^{i\theta_0}$ and $\Delta(\pm\mathbf Q^*)=|\Delta_{\mathrm{PDW}}|e^{i\theta\pm i\phi}$.
The corresponding real-space pairing profile is
\begin{equation}
\begin{aligned}
\Delta(\mathbf R)={}|\Delta_{\mathrm{SC}}|e^{i\theta_0}+2|\Delta_{\mathrm{PDW}}|e^{i\theta}
\cos(\mathbf Q^*\cdot\mathbf R+\phi),
\end{aligned}
\label{eq:sc_pdw_parametrization}
\end{equation}
where $\mathbf R$ is the pair center-of-mass coordinate, $\phi$ specifies the position of the modulation, and $\theta-\theta_0$ is the relative phase between the PDW and SC components.
Our numerical results are consistent with $|\Delta_{\mathrm{PDW}}|\approx\tfrac{1}{2}|\Delta_{\mathrm{SC}}|$ and $\theta-\theta_0\approx0$, giving $\Delta(\mathbf R)\propto[1+\cos(\mathbf Q^*\cdot\mathbf R+\phi)]$.
This agrees well with the numerical real-space pairing profile in Fig.~\ref{fig:NonScanCylinder}(e), whose period of approximately seven lattice spacings along the cylinder is consistent with $|Q_x^*|=2\pi/7$.

The flat edges of the Fermi surface thus appear to favor finite-momentum pairing in this model at higher doping, in contrast to the uniform SC phase that emerges from a nearly isotropic Fermi surface at lower doping.
However, the observed wavevector $\mathbf Q$ appears tied to the discrete $k_y$ values imposed by the finite cylinder width, leaving unresolved whether this mechanism would still favor PDW in the two-dimensional thermodynamic limit and, if so, which wavevector it would select.

{\it Summary.}---%
We have shown that doping a single repulsive Dirac fermion produces $p-ip$ superconductivity. This behavior seems to extend down to very low doping and emerges from a single, nearly isotropic Fermi surface, without fine tuning of its shape or additional longer-range hopping.
We hope these results motivate further theoretical work on the pairing mechanism in this model. The nearly complete orbital polarization we observe may allow for an effective weakly-interacting description~\cite{DongLee2025} (see also Ref.~\cite{DongLeeTriangular2025}), although the fact that the Dirac mass is dynamically generated by the repulsion could complicate the analysis.

\begin{acknowledgments}
{\it Acknowledgments.}---%
Work by T.S. and O.T. was supported by the U.S. Department of Energy, Office of Science, Office of Basic Energy Sciences under Early Career Research Program Award No. DE-SC0025568. 
CAG was supported by the UCI-LANL-SoCal Hub Graduate Fellowship program and by the Eddleman Quantum Institute at UCI.
CY was supported by the Eddleman Quantum Institute postdoctoral fellowship and the Moore Foundation postdoctoral fellowship. 
SRW was supported by the U.S. NSF under Grant DMR-2412638. 
We gratefully acknowledge discussions with Federico Becca, Sasha Chernyshev, Luis Jauregui, Patrick Lee, Subir Sachdev, Javier Sanchez-Yamagishi, Shengtao Jiang, and Miguel Gon\c{c}alves.
The use of AI tools, specifically OpenAI’s ChatGPT and Codex (Sol and Astra), was limited to proofreading the draft to improve grammar and style and assisting with figure preparation.
\end{acknowledgments}

\bibliography{p-ip_DMRG}
\clearpage

\onecolumngrid
\begin{center}
\ \vskip 0.2cm
{\large\bf End Matter}
\end{center}
\twocolumngrid

\renewcommand{\theequation}{A\arabic{equation}}
\setcounter{equation}{0}

{\it Appendix A: DMRG details}.---%
For the first scan in Fig.~\ref{fig:Scan_INS-SC}, no pairing field is applied. 
To check convergence, we use several initial states, including an orbital-polarized insulator and a N\'eel state. 
In all cases, pairing develops within the first few DMRG sweeps, and the calculations converge to the same final paired state.

For the wider scan in Fig.~\ref{fig:Scan_SC-PDW}, convergence is more challenging. 
We initialize the calculation from a state obtained in a narrower low-doping scan and apply a homogeneous nearest-neighbor pinning pair field of strength $10^{-4}$ in the $p-ip$ sector; an approach that has proven effective in studies of SC phases~\cite{ttprimej, tJsteve1998, HubbardPRX2020, Chung2020, DownfoldingHubbard2023, HubbardScience2024}. 
Testing different pairing symmetries with the pinning pair field, we find a strong response only in the $p-ip$ channel.
The field also fixes the phase of the SC order parameter and prevents DMRG from restoring $U(1)$ symmetry, which would otherwise cause the local pair amplitude to vanish.

{\it Appendix B: Momentum distribution $n({\bf k})$.}---%
The momentum distribution $n({\bf k})=\sum_{\alpha={\rm A,B}} \langle c_{{\bf k},\alpha}^\dagger c_{{\bf k},\alpha}\rangle$ reveals a relatively sharp Fermi surface in the SC phase [Figs.~\ref{fig:1RDM_Nk}(b) and (d), plotted as the hole distribution $\delta_{\bf k}=1-n({\bf k})$].
Its shape and size agree qualitatively with the Fermi surface obtained from the model introduced in the main text at the same hole doping, with Hartree mass $m_{\rm eff}=U/2$ (dashed contours).
As doping increases from $\delta\approx0.02$ to $\delta\approx0.1$, the hole pocket expands to intersect the $k_y=\pm\pi/4$ channels in addition to $k_y=0$, consistent with the crossover from quasi-1D to isotropic SC described in the main text.
The sorted eigenvalues of the one-body reduced density matrix [Figs.~\ref{fig:1RDM_Nk}(a) and (c)] likewise show a broadened step in the $k_y=0$ sector at both dopings, with additional steps in the $k_y=\pm\pi/4$ sectors at $\delta\approx0.1$.

{\it Appendix C: Pairing at finite total momentum.}---%
As mentioned in the main text, in the PDW+SC phase additional finite-momentum components develop, most prominently at ${\bf Q}=\pm{\bf Q}^*$, where ${\bf Q}^*=(-2\pi/7,\pi/4)$.
Figure~\ref{fig:DeltaQ} shows the ${\bf k}$ dependence of $\Delta_{{\bf k},{\pm {\bf Q}^*}}$ for the same PDW+SC state as in Fig.~\ref{fig:Pairing}(f).

\newpage 
\ \
\newpage 
\ \
\newpage
\onecolumngrid
\begin{center}
{\large\bf Repulsion-Driven $p - i p$ Superconductivity in a Single Valley Revealed by DMRG \\ Supplemental Material}\\ 
\vskip0.35cm
Cesar A. Gallegos$^1$, Omid Tavakol$^1$, Christopher Yang$^1$, Steven R. White$^1$, and Thomas Scaffidi$^1$\\
\vskip0.15cm
{\it \small $^1$Department of Physics and Astronomy, University of California, Irvine, California
92697, USA}\\
{\small (Dated: \today)}\\
\end{center}
\vskip -0.5cm \

\setcounter{page}{1}
\thispagestyle{empty}
\makeatletter
\renewcommand{\c@secnumdepth}{0}
\makeatother
\setcounter{section}{0}

\section{Hartree Mean-Field Calculation}

In this section, we describe how the on-site repulsion $H_U$ generates the effective mass $m_{\mathrm{eff}}$ discussed in the main text.
We denote the average electronic density as $n_{\mathrm{tot}}=n_{\rm A}+n_{\rm B}$ and the average orbital occupations by $n_\alpha=\langle n_{i,\alpha}\rangle$, with
$\alpha={\rm A},{\rm B}$.

Using a mean-field approximation, the on-site interaction
\begin{equation}
H_U=U\sum_i n_{i,{\rm A}}n_{i,{\rm B}}
\end{equation}
is decoupled as
\begin{equation}
n_{i,{\rm A}}n_{i,{\rm B}}
\simeq n_{\rm B} n_{i,{\rm A}}+n_{\rm A} n_{i,{\rm B}}-n_{\rm A} n_{\rm B}.
\end{equation}
In terms of the orbital spinor $c_i=(c_{i,{\rm A}},c_{i,{\rm B}})^T$, this gives the Hartree approximation
\begin{equation}
H_U^{\mathrm H}
=
\sum_i c_i^\dagger
\left[
\frac{U n_{\mathrm{tot}}}{2}\mathbb{1}
-\frac{U(n_{\rm A}-n_{\rm B})}{2}\sigma^z
\right]c_i
-N U n_{\rm A} n_{\rm B},
\end{equation}
where $N$ is the number of lattice sites. 
This adds a uniform chemical potential shift $\mu_{\mathrm{eff}} \mathbb{1}$ and a mass term $m_{\mathrm{eff}}\sigma^z$ to the single-particle Hamiltonian, where
\begin{equation}
\mu_{\mathrm{eff}} = \frac{U n_{\mathrm{tot}}}{2} , \quad \mathrm{and} \quad  m_{\mathrm{eff}}=-\frac{U(n_{\rm A}-n_{\rm B})}{2}.
\label{eq:hartree_parameters}
\end{equation}

Substituting the Hartree approximation $H_U \to H_U^H$ into the Hamiltonian $H = H_0 + H_U$, the Hartree Hamiltonian in the continuum limit is given by
\begin{equation}
\mathcal H_{\mathrm H}(\bk)
=
v\sin k_y\,\sigma^x-v\sin k_x\,\sigma^y
+d_z(\bk)\sigma^z + \mu_{\mathrm{eff}} \mathbb{1},
\end{equation}
with
\begin{equation}
d_z(\bk)=m_{\mathrm{eff}}
+\mathcal{B}(2-\cos k_x-\cos k_y).
\end{equation}
Near the $\Gamma$ point of the Brillouin zone, it can be expanded as
\begin{equation}
\mathcal H_{\mathrm H}(\bk)
\simeq
v k_y\sigma^x-v k_x\sigma^y
+\left(m_{\mathrm{eff}}+\frac{\mathcal{B}}{2}k^2\right)\sigma^z  + \mu_{\mathrm{eff}} \mathbb{1}.
\end{equation}
Thus, the Hartree approximation predicts the opening of a mean field gap with size $2 |m_{\mathrm{eff}}|$ at the $\Gamma$ point.

To calculate $m_{\mathrm{eff}}$ and $\mu_{\mathrm{eff}}$, the orbital occupations $n_{\rm A}$ and $n_{\rm B}$ must be determined self-consistently from the band dispersion of $\mathcal H_{\mathrm H}(\bk)$. 
The band energies are given by
\begin{equation}
E_{\bk,\nu}=\nu\epsilon_{\bk}-\mu_{\mathrm{eff}},
\qquad
\epsilon_{\bk}
=
\sqrt{v^2(\sin^2 k_x+\sin^2 k_y)+d_z(\bk)^2},
\qquad \nu=\pm.
\end{equation}
Using $n_\alpha=\langle c_{i,\alpha}^\dagger c_{i,\alpha}\rangle$, we perform a Fourier transform to obtain
\begin{equation}
n_\alpha
=
\frac{1}{N}\sum_{\bk}
\langle c_{\bk,\alpha}^\dagger c_{\bk,\alpha}\rangle,
\end{equation}
where $c_{i,\alpha} = \frac{1}{\sqrt{N}}\sum_{\bk} e^{i\bk\cdot\mathbf{r}_i}\,c_{\bk,\alpha}$, $\mathbf{r}_i$ is the spatial coordinate of site $i$, and $N$ is the number of lattice sites. 
Transforming to the eigenbasis of the Hartree Hamiltonian,
\begin{equation}
c_{\bk,\alpha}
=
\sum_{\nu=\pm}u_{\alpha\nu}(\bk)\gamma_{\bk,\nu},
\end{equation}
where $\gamma_{\bk,\nu}$ annihilates a fermion of momentum $\bk$ in band $\nu$ with energy $E_{\bk,\nu}$.
In thermal equilibrium, $\langle\gamma_{\bk,\nu}^\dagger\gamma_{\bk,\nu'}\rangle = \delta_{\nu\nu'}n_{\mathrm F}(E_{\bk,\nu})$, where $n_F(E)$ is the Fermi-Dirac distribution, so
\begin{equation}
n_\alpha
=
\frac{1}{N}\sum_{\bk,\nu}
|u_{\alpha\nu}(\bk)|^2 n_{\mathrm F}(E_{\bk,\nu}).
\end{equation}

Using $|u_{{\rm A}\nu}(\bk)|^2+|u_{{\rm B}\nu}(\bk)|^2=1,$ and 
\begin{equation}
|u_{{\rm A}\nu}(\bk)|^2-|u_{{\rm B}\nu}(\bk)|^2
=
\nu\frac{d_z(\bk)}{\epsilon_{\bk}},
\end{equation}
we obtain
\begin{equation}
n_{\mathrm{tot}}
=
\int_{\mathrm{BZ}}\frac{d^2k}{(2\pi)^2}
\sum_{\nu=\pm}n_{\mathrm F}(E_{\bk,\nu}),
\end{equation}
\begin{equation}
n_{\rm A}-n_{\rm B}
=
\int_{\mathrm{BZ}}\frac{d^2k}{(2\pi)^2}
\sum_{\nu=\pm}
\nu\frac{d_z(\bk)}{\epsilon_{\bk}}
n_{\mathrm F}(E_{\bk,\nu}).
\end{equation}
These equations close the
Hartree self-consistency loop: the occupations determine $\mu_{\mathrm{eff}} = Un_{\mathrm{tot}} / 2$ and $m_{\mathrm{eff}}=U(n_{\rm B}-n_{\rm A})/2$, which in turn determines the band energies and orbital polarization $d_z(\bk)$.

We will focus on the light doping limit near half-filling. 
At half filling, the Fermi-Dirac distribution is given by $n_F(E_{\bk,\nu}) = \delta_{\nu, -}$ at zero temperature, and thus
\begin{equation}
n_{\rm A}-n_{\rm B}
=
-\int_{\mathrm{BZ}}\frac{d^2k}{(2\pi)^2}
\frac{d_z(\bk)}{\epsilon_{\bk}}.
\end{equation}
In the non-interacting limit $U=0$ and for $\mathcal{B}>0$, $d_z(\bk)=\mathcal{B}(2-\cos k_x-\cos k_y)$ is nonnegative throughout the Brillouin zone. 
The occupied non-interacting band therefore has larger weight in the B orbital, $n_{\rm B}>n_{\rm A}$. 
Performing the self-consistent calculation iteratively leads to an enlarged orbital polarization $d_z(\bk)\to m_{\mathrm{eff}} + \mathcal{B}(2-\cos k_x-\cos k_y)$, because $m_{\mathrm{eff}}=U(n_{\rm B}-n_{\rm A})/2$ is positive. 
The final self-consistent solution in the large $U$ limit is an occupied band nearly fully polarized in the B orbital, giving $n_{\rm B}\simeq1$ and $n_{\rm A}\simeq0$, so that $m_{\mathrm{eff}}\simeq U/2$. 
Straightforwardly generalizing the calculation to $\mathcal{B}<0$, we obtain the generic result
\begin{equation} \label{eq:meff}
m_{\mathrm{eff}}\simeq\frac{U}{2}\operatorname{sgn}(\mathcal{B}).
\end{equation}
The Hartree mass therefore has the same sign as the Wilson term, yielding the topologically trivial insulating state described in the main text. 

\begin{figure}[h!]
\includegraphics[width=0.4\linewidth]{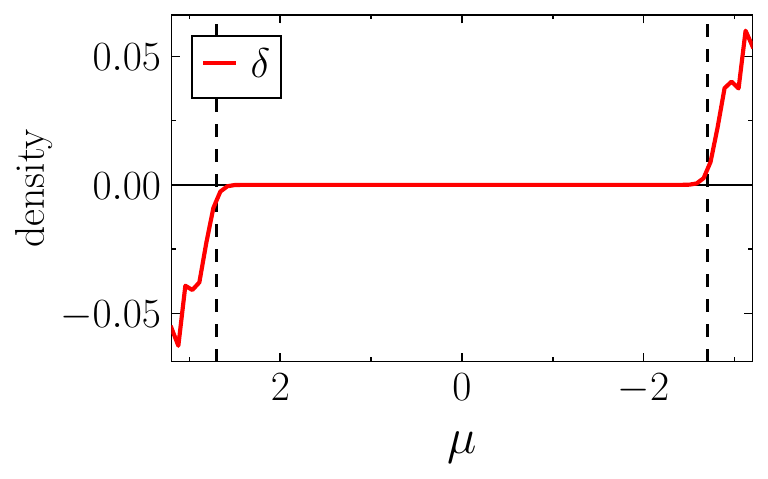}
\vskip -0.2cm
\caption{Total hole density $\delta = 1 - n_{\mathrm{A}}- n_{\mathrm{B}}$ vs chemical potential $\mu$ in a DMRG scan on an $84 \times 8$ cylinder for $U = 6$ and $\mathcal{B} = 0.5$, with the chemical potential decreasing linearly from $3.2$ to $-3.2$. Here the wide scan displays the entire insulating gap, with the sharp changes in $\delta$, indicated by the dashed lines, corresponding to the band edge of the conduction and valence bands. The corresponding gap size is estimated to be $5.4$.}
\label{fig:ph_scan}
\end{figure}

To compare the predictions of mean field theory to DMRG, we note that the observed orbital polarization at $U = 6$ in the insulating gap is imperfect, see Fig.~\ref{fig:1RDM_Nk}(d), with $n_A \approx 1-n_B \approx 0.02$. 
Thus, by Eq.~\eqref{eq:meff}, we would expect to find a gap of size $2|m_{\mathrm{eff}}| \approx 5.76$. 
This is larger than the gap measured from a wide DMRG scan, see Fig.~\ref{fig:ph_scan}, which estimates its size to be $5.4$. 
This is an indication that there are non-trivial correlation effects beyond mean-field theory.

\section{Width-6 scan}
\begin{figure}[h!]
\includegraphics[width=0.9\linewidth]{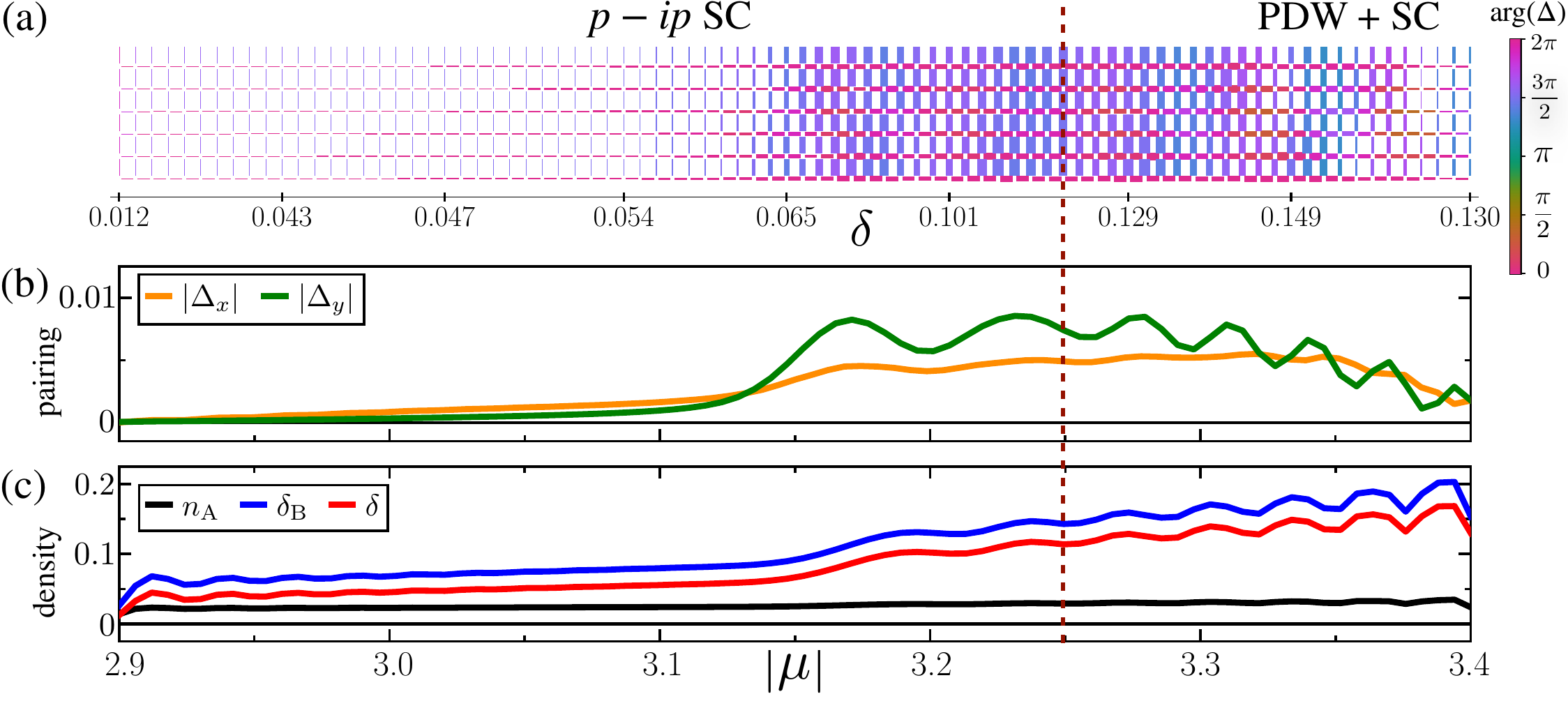}
\vskip -0.2cm
\caption{Same as Fig.~\ref{fig:Scan_SC-PDW} of the main text, but in a width-6 cylinder.}
\label{fig:W6Scan}
\end{figure}

In this section, we present the same results as in Fig.~\ref{fig:Scan_SC-PDW}, but for width-$6$ cylinders. 
Fig.~\ref{fig:W6Scan} shows that the quasi-1D-to-isotropic transition occurs at $\delta\approx0.065$, a higher doping than in the width-$8$ cylinder shown in the main text. 
This shift is consistent with the expected $2/N_y^2$ dependence of the transition point on cylinder width, further supporting the interpretation of the quasi-1D SC behavior as a finite-width effect.

\end{document}